\documentstyle[epsf]{l-aa}     
\begin{document}
   \thesaurus{07         
              (07.09.1;  
               03.01.1;  
               16.01.1;  
               04.01.1)  
             }
   \title{First DENIS I-band extragalactic catalog \thanks{Based on observations collected at the European Southern Observatory, La Silla, Chile.}}

   \author{Vauglin I., \inst{1} Paturel G., \inst{1} 
 Borsenberger J., \inst{2}  Fouqu\'e P., \inst{3,4}
 Epchtein N., \inst{5,3}  Kimeswenger S.,\inst{6} 
 Tiph\`ene D.\inst{3} Lanoix P.,\inst{1} Courtois H., \inst{1} }
 
    \offprints{G. Paturel}
 
   \institute{
              CRAL-Observatoire de Lyon,\\
              F69561 Saint-Genis Laval CEDEX, FRANCE \\
\and
              Institut d'Astrophysique de Paris\\
              98 bis boulevard Arago, F75014 Paris, FRANCE \\
\and
              Observatoire de Paris-Meudon\\
              F92195 Meudon Principal CEDEX, FRANCE\\
\and
              European Southern Observatory\\
              La Silla, La Serena, CHILE\\
\and
              Observatoire de Nice\\
              Departement Fresnel, BP 4429, F06304 Nice CEDEX, FRANCE\\
\and 
              Institut fur Astronomie\\
              Technikerstrasse 25, A6020 Innsbruck, AUSTRIA\\
}
 
   \date{Submitted August, 1998; Accepted            }
 
   \maketitle
   \markboth{Vauglin I. et Al.} {DENIS extragalactic catalog}

   \begin{abstract}
This paper presents the first I-band photometric catalog of the brightest
galaxies extracted from the Deep Near Infrared Survey of the Southern Sky (DENIS)
An automatic galaxy recognition program has been developed to build
this provisional catalog. 
The method is based on a discriminating analysis.
The most discriminant parameter to separate galaxies from stars is proved to be
the peak intensity of an object divided by its array. 
Its efficiency is better than 99\%.
The nominal accuracy for galaxy coordinates calculated with the Guide Star Catalog
is about 6 arcseconds.
The cross-identification with galaxies available in the Lyon-Meudon
Extragalactic DAtabase (LEDA)
allows a calibraton of the I-band photometry with the sample of
Mathewson et Al.
Thus, the catalog contains total I-band magnitude, isophotal diameter,
axis ratio, position angle and a rough estimate of the morphological type code
for 20260 galaxies. The internal completeness of this catalog reaches 
magnitude $I_{lim}=14.5$, with a photometric accuracy of $\sim 0.18m$.
25\% of the Southern sky has been processed in this study.

This quick look analysis allows us to start a radio and spectrographic 
follow-up long before the end of the survey.

   \end{abstract}
\section{Introduction} 
In Paturel et al. (1996),
we presented our program of collecting the main astrophysical parameters for
the principal galaxies. 
The first target was limited to adding information on galaxies
already known in the LEDA database. 

The work is now more ambitious because we are aiming at detecting
new galaxies from the {\it Deep Near Infrared Survey of the Southern
Sky} (hereafter DENIS). 
DENIS is a program to survey the entire southern sky in three wavelength bands
(Gunn-i: 0.80$\mu m$, J: 1.25$\mu m$ and Ks: 2.15$\mu m$) with 
limiting magnitudes
of 18.5, 16.5 and 14.0 respectively. The observations are 
performed with the ESO 1m-telescope at La Silla (Chile) with a dedicated camera.
The survey observations with three channels started in routine mode in December
1995. A detailed description of DENIS is given in Epchtein (1998) and
in Garz\'on et al. (1997). 

The systematic detection, extraction and cataloging of DENIS extragalactic sources
are of significant interest for studies requiring large and homogeneous samples
such as the kinematics of the local universe, the distance scale, 
cosmology etc...
The I-band is the most suitable both for the detection of extended 
objects and for the star/galaxy separation, except in the galactic plane.
Thus, in the present study only the I-band images are considered. 

The transfer of DENIS images from Paris to Lyon is explained in section 2 and
the extraction of objects from these images in section 3.
In sections 4 to 5 we describe astrometry and automatic galaxy 
recognition and analysis. Then,
in sections 6 to 8, we explain how galaxies are cross-identified with LEDA galaxies
leading to the comparison of astrophysical parameters with those
from Mathewson et al. and from LEDA.
Finally, in section 9 we describe the provisional I-band DENIS catalog. 

It is important to note that a deeper catalog will be made at the end
of the survey. Hence, the present catalog is a preliminary catalog
of bright galaxiesdetected by DENIS, and
used to start a radio and spectrographic follow-up long
before the end of the survey.
The present measurements cover one year of observation.

\section{Obtaining I-band CCD frames from DENIS}
\subsection{Characteristics of the DENIS survey}
The I-band of the survey is the {\it Gunn-i} band
at $0.80 \mu m$. The CCD camera  is a Tektronix 1024$\times$1024 pixel array cooled 
down to liquid nitrogen temperature.
Each frame (768$\times$768 pixels) represents $12$ square arcminutes with a pixel size of
$1$ arcsecond. The integration time is $9$ seconds. The read-out noise is about
$7 e^{-}$. 
The observing strategy consists in scanning at a constant right ascension on
strips of $30$ degrees in declination (180 frames per strip) taken in three zones
``Equatorial'' from $\delta=+2 \deg$ to $-28\deg$, ``Intermediate'' from $\delta=-28 \deg$ to $-58\deg$
and ``Polar'' from $\delta=-58 \deg$ to $-88\deg$. The overlap between 
adjacent frames
is $1$ arcminute on each side (i.e. $2$ arcminute in each direction). 
This strategy aims at covering a wide range of airmasses.
Each strip starts and ends with photometric and astrometric calibrations. At the end
of some nights a flat fielding is performed directly on the sky during sunrise. The data
is archived on DAT cartridges which are send each week to the {\it Paris
Data Analysis Center} (PDAC) at the Institut 
d'Astrophysique de Paris for processing.
\subsection{Pipeline Lyon-PDAC}
\begin{figure}
\epsfxsize=8.5cm
\hbox{\epsfbox{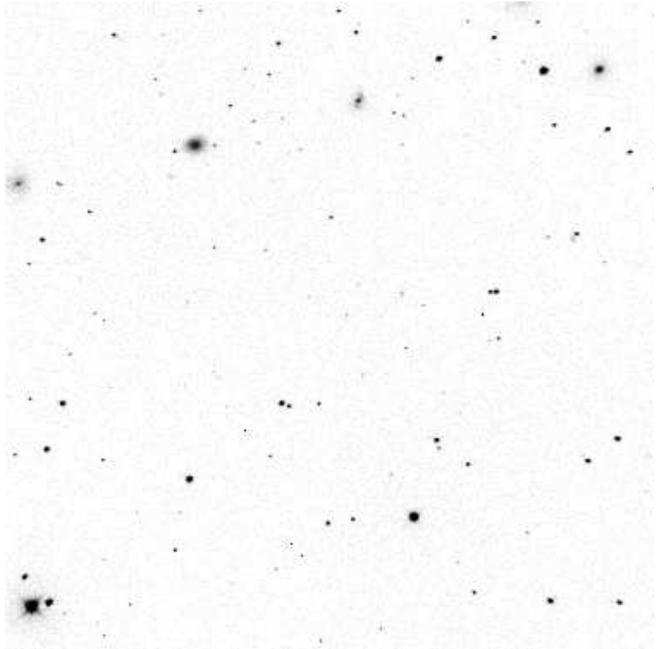}}
\caption{A typical image from the DENIS Survey after the $2 \times 2$ 
rebinning. The frame is
$12' \times 12'$. North is on the bottom side, East is on the left side.
Among the four galaxies clearly visible on the top, one (the third from the left) is a new one.
This shows that relatively bright galaxies can be discovered.}
\label{fig1}
\end{figure}

A systematic automatic processing of DENIS I-images began in Lyon
in February 1996.
Flat-fielding and de-biasing are made at PDAC on each genuine 
frame. For the Lyons
processing each I-image is reduced by a factor 4 in size by
rebinning pixels 2$\times$2. Our effective pixel size is thus $2$ arcseconds.
An example of an I-frame is given in Figure~\ref{fig1}.

The histogram of pixel intensities is used to calculate
the sky background intensity $f_{bg}$.
The sky-background level is taken at the maximum intensity of the histogram  
(i.e. the mode)
and the standard deviation $\sigma$ is calculated by symmetrizing the 
low intensity part of the histogram with respect to the mode. 
The sources are concentrated in the high intensity part of the
histogram.

A threshold is then applied (the threshold level is $f_{bg} + 2.0 \sigma$).
Using this procedure (averaging and thresholding) allows a compression factor of
20 to 30, depending on the image contents. 
All images of a given I-strip are thus compressed, tar'd and automatically
transferred to Lyon via ftp. A full strip is stored in 10 to 13 Mbytes.
The galaxy extraction is made at Lyon using the program described in the
following section.

\section{Extraction of astronomical objects}
\begin{figure}
\epsfxsize=5.5cm
\hbox{\epsfbox{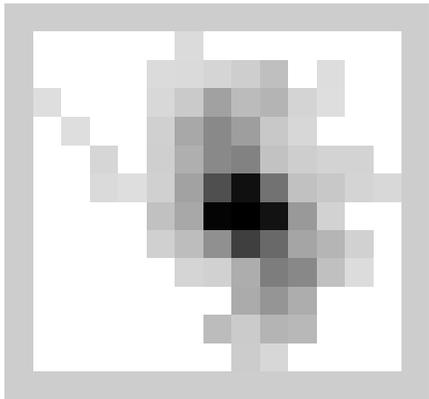}}
\caption{Example of a matrix for a small galaxy. One pixel is 
$2\prime\prime \times 2\prime\prime$. The edge
is outside the matrix.}
\label{fig3}
\end{figure}

All sources (stars, galaxies, defects etc...) are extracted using the same
algorithm as described in Paturel et al. (1996, section 3.1),
except that no attempt was made to share interacting objects which
are simply flagged after visual inspection (section 5). The reason is
that we are interested first of all in well defined objects. 
At the end of this stage, we obtain for each frame a collection of matrixes
(see an example in Figure \ref{fig3}). Matrixes smaller than 17 pixels are rejected. 
They correspond, to the mean, of objects of $6$ arcseconds in  diameter. 

Astrophysical parameters are extracted for each matrix according to Paturel et 
al. (1996, section 3.2). These parameters are the following:

\begin{itemize}
\item Weighted mean position of the center $x_m$ and $y_m$ (in pixels),
where the weight is the pixel intensity $f_{ij}$
\item major axis $logD$ (D in $0.1 arcmin.$) at faint isophotal level (external diameter).
\item axis ratio $logR$  (log of major to minor axis)
\item position angle of the major axis $\beta$ (in degrees, counted from
North towards East)
\item magnitude $I= -2.5 log (\sum_{i=1}^{n} f_{ij}- f_{bg}) + cst$ (in arbitrary units), 
$n$ is the number of pixels with intensity $f_{ij}$ larger than the threshold.
\end{itemize}

 We have now to perform astrometry (conversion  of pixels positions to 
right ascension and declination) and then  recognition
of ``galaxies'', ``stars'', and ``unknown objects''. 

\section{Astrometry}
\begin{figure}
\epsfxsize=8.5cm
\hbox{\epsfbox{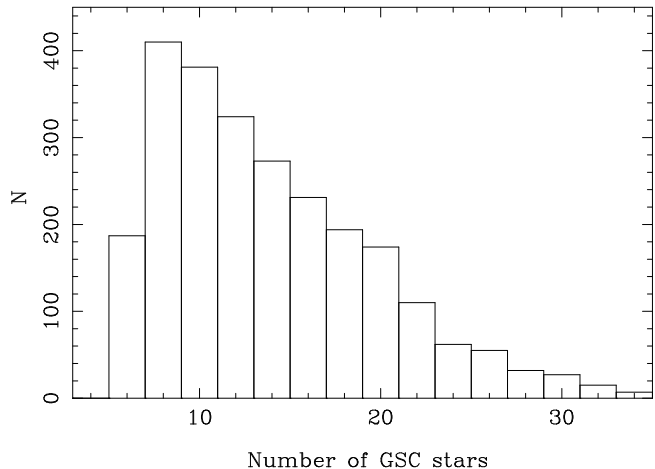}}
\caption{Histogram of the number of GSC stars by frame.} 
\label{fig4}
\end{figure}

The center of the frame is taken from the header of the FITS file.
From the Guide Star Catalog (GSC)
and from the LEDA database we extract all objects (stars or 
galaxies) known in the corresponding square.
A cross-identification between matrixes and stars is made exactly 
as described in Paturel et al. (1996, sections 3.3 and 3.4).
Galaxies are also used but only when they have accurate coordinates 
(i.e. typically better than $10$ arcsecond). A 6th order polynomial fit
converts (x,y) positions on the frames
to Right Ascension and Declination.
The number of GSC stars varies from one frame to another. A histogram
of number of GSC stars per frame is given in Figure~\ref{fig4}. If this number is
smaller than 7 or if the standard deviation of the polynomial fit is
greater than $4$ arcseconds, the solution is rejected and we adopt 
the 'header' solution calculated from the coordinates of the center and
the pixel size as given in the header. 
If the GSC-solution seems acceptable but differs from the header-solution
by more than $30$ arcseconds, the header solution is preferred and 
coordinates are flagged to recall that they may be inaccurate.

\begin{figure}
\epsfxsize=8.5cm
\hbox{\epsfbox{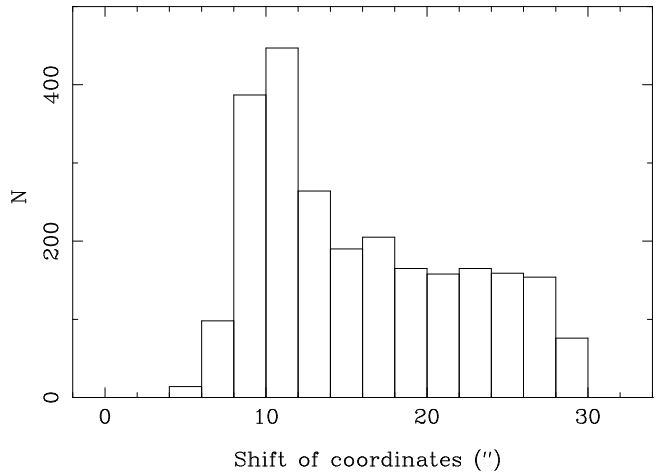}}
\caption{Histogram of the differences between the two astrometric solutions
(GSC- and direct-solution).} 
\label{fig5}
\end{figure}

In Figure \ref{fig5}, we show the differences between the GSC-solution and the
FITS header-solution. Most of them are in good agreement within $15$ arcseconds.
Note that more recent measurements have been astrometrically calibrated to
better than 1 arcsecond by cross-identifying with the PMM database.

Galaxy coordinates will be compared directly with coordinates of LEDA galaxies.

\section{Automatic star/galaxy separation}
\subsection{Test sample}
A sample of 1146 objects has been visually classified into 
three classes: stars, galaxies and unknown objects. The
distribution in each class is the following:

\vspace{0.5cm}
\begin{tabular}{lrr}
\hline
Class& Number of objects  & percentage \\
\hline
Stars     & 500   & 43.6\% \\
Galaxies  & 523   & 45.6\% \\
Unknown   & 123   & 10.7\% \\
\hline
Total     &1146   &  100\% \\
\hline
\end{tabular}
\vspace{0.5cm}

This sample will be used as a test sample (or training sample) in a 
discriminant analysis method (DA) for galaxy recognition.

\subsection{Discriminant analysis}
The DA is a common method for automatic recognition. 
The test sample being shared in $n_c$ classes $\{I_k\}$ ($k=1,n_c$),
the purpose of DA is to find the principal factorial
axis on which each class is as concentrated as possible
and as distinct as possible from the others. This is achieved by maximizing
the inertia between classes and by minimizing the inertia within each class.
Inertia is calculated from the set of parameters attached to each object.
We will note $p_{ij}$ the $j$-th parameter of object $i$. 

The mathematical result (see Diday et al., 1982) is that the factorial axes 
are the eigenvectors of the matrix ${\cal{T}}^{-1}.\cal{B}$, where
$\cal{T}$ is the total covariance matrix and $\cal{B}$ is the
inter-class covariance matrix.

Note that the matrix  ${\cal{T}}^{-1}.\cal{B}$ is not symmetrical and that
the total covariance matrix is the sum of
the intra-class covariance matrix $\cal{W}$ (covariance Within class) 
and of inter-class covariance matrix $\cal{B}$ (covariance Between class).
This is called the Huyghens decomposition.  
\begin{equation}
\cal{T}= \cal{W} + \cal{B}
\end{equation}

The elements of $\cal{B}$ are:
\begin{equation}
b_{jj'} = \sum_{k=1}^{n_c} \frac {N_k}{N} (\bar{p_{j}^{k}} - \bar{p_j})(\bar{p_{j'}^{k}} - \bar{p_{j'}})
\end{equation}
where $N_k$ is the number of objects in the class $I_k$  (k=1 to $n_c$), where the mean
parameter $j$ for the whole sample is:
\begin{equation}
\bar{p_j}= \frac {1}{N} \sum_{i}p_{ij}
\end{equation}
(N is the number of objects of the whole sample), and where the mean parameter $j$ 
within the class $\{I_k\}$ is:
\begin{equation}
\bar{p_{j}^{k}}= \frac {1}{N_k} \sum_{i \in I_k}p_{ij}
\end{equation}
The elements of the total covariance matrix $\cal{T}$ are:
\begin{equation}
t_{jj'} = \sum_{k} \frac {1}{N}  \sum_{i \in I_k} (p_{ij} - \bar{p_j}) (p_{ij'} - \bar{p_{j'}})
\end{equation}

Now, we have to choose the set of parameters attached to each object. 

\subsection{Choice of discriminant parameters}
Any discriminant method requires a good choice of discriminant 
parameters which are used for the definition of the metric.  
These parameters are not necessarily independent but they must 
cover all features which seem relevant for a reliable discrimination of 
astronomical objects. For galaxy recognition we tested 7 parameters.
\begin{enumerate}
\item  Peak intensity per area unit, this is Peak intensity divided by the
surface of the considered object. 
\item  Mean surface brightness, total flux divided by area
\item  Peak intensity,
\item  Axis ratio, ratio of the major to the minor axis 
\item  Relative area, ratio of number of pixels of the object and of the matrix.
\item  Elongation of the matrix 
\item  Presence of diffraction cross
\end{enumerate}
The DA method is applied on half the sample (i.e. 573 objects) and
tested on the other half using only one parameter at a time (in this case
the factorial axis is defined by the parameter itself). 
The percentage of good results is given below for
each one, individually.

\vspace{0.5cm}
\begin{tabular}{lrrr}
\hline
Parameter            &  stars    &  galaxies   &   mean\\
\hline
Peak over area       &  100.0\%  &    98.9\%   &   99.4\%         \\
Mean SB              &   98.7\%  &    99.6\%   &   99.2\%         \\
Peak intensity       &   96.2\%  &    98.5\%   &   97.4\%         \\
Axis ratio           &   83.7\%  &    54.8\%   &   69.2\%         \\
Relative surface     &   77.0\%  &    58.5\%   &   67.7\%         \\
Elongation of matrix &   88.3\%  &    40.8\%   &   64.5\%         \\
Diffraction Cross    &   68.2\%  &    57.0\%   &   62.6\%         \\
\hline
\end{tabular}
\vspace{0.5cm}

The conclusion of this test is that the most relevant information about
the nature of an object is contained in the pixel intensity, not in the shape of the object.
Stars have a very high central intensity, galaxies do not. Moreover, stars
are concentrated, galaxies are not.
This explains why ``Mean SB'' and ``Peak over area'' give such an impressive recognition rate.
Finally, only the first four parameters have been used.
The axis ratio is kept because it becomes relevant for faint objects despite
that its rate is relatively low.

\subsection{Result}
\begin{figure}
\epsfxsize=8.5cm
\hbox{\epsfbox{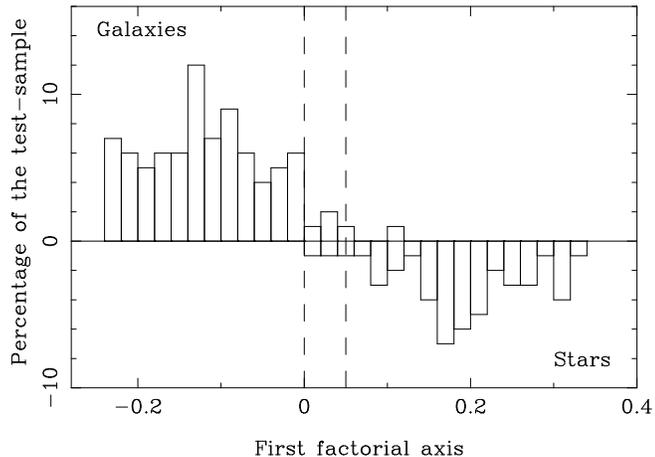}}
\caption{Definition of acceptation zones along the first factorial axis.
The left-hand zone defines ``Galaxies'', the righthand one defines ``Stars'',
and the intermediate one defines ``undefined objects''.}
\label{fig6}
\end{figure}

The DA method is applied with the four parameters described above and three
classes ``Galaxies'', ``Stars'' and ``unknown objects''. 
Using the test sample, each object
is projected onto the first factorial axis. Figure \ref{fig6} shows
the projection onto the first factorial axis of ``Galaxies'' and 
``Stars'' classes. Similar plots exists for ``Stars'' and ``unknown
objects'' classes and for ``Galaxies'' and ``unknown objects''
classes. All ``unknown
objects'' have been eliminated in the next part of this study. \\
One can see that there is an overlapping  region where ``Galaxies''
and ``Stars'' are mixed. The limits of this zone can be tuned in such
a way that one can accept a given percentage of misclassification. We choose
$0\%$ chance of classifying a star as a galaxy and $5\%$ chance of classifying
a galaxy as a star. Indeed, it is important to avoid the contamination
of the catalog by stars while it is not as important to miss a galaxy (which
is uncertain anyway). These limits are drawn on Figure \ref{fig6} where it is
visible that no star enter the galaxy-domain, while $5\%$ of galaxies enter
the star-domain. Objects between these two limits will be classified 
as undefined.

\subsection{Visual control}
The final step of this treatment consists in checking visually all frames
recognized as galaxies. This tedious part allows us to reject artefacts 
(1148 rejections after the inspection of 54073 images)
like those produced by star halos truncated by the edge of the frame.
Such truncated halos look like elongated, low-surface brightness object,
easily accepted as galaxies. 

As a result, a code is given to describe three features:
\begin{itemize}
\item ``multiple'', if several objects are present in the matrix
\item ``truncated'', if the galaxy is truncated by the edge of the array.
\item ``peculiar'', if the galaxy looks strange for any reason
\end{itemize}
So, each galaxy of the catalog has been inspected visually. This will prevent
us from gross misidentification. 
Now, the galaxies have to be cross-identified with known galaxies. 

\section{Cross-identification}
To make a correct cross-identification we need
coordinates, calibrated magnitudes and diameters, axis
ratios and position angles. 
We will use the first calibration obtained from the
preliminary cross-identification made for astrometric purposes (section 4).
This calibration will be refined in next section.

Because of frame overlap along the strip, many objects are measured twice 
(or even three or four times with adjacent strips).
A first cross-identification is done for these galaxies measured several times. 
This will be called the ``auto-cross-identification''. 
Then, the cross-identification with LEDA galaxies may start.

\subsection{Auto-cross-identification}
The ``auto-cross-identification'' is performed using a hierarchical method in which 
we merge step by step the closest objects. The definition of the distance of two 
objects $i$ and $j$ is the following:
\begin{equation}
d_{ij} = \frac{1}{n_p} \sum_{k=1}^{n_p} \frac {|p_{ik} - p_{jk}|}{2 \sigma_k}
\label{distance}
\end{equation}
where, $n_p$ is the number of parameters (coordinates, diameters magnitudes...) for
a given object,
$p_{ik}$ is the k-th parameter of object $i$,
$\sigma_k$ is the standard error of the $k-th$ parameter.
When two objects are merged they are replaced by a single one, whose parameters
are the means of both. The final result does not depend on the order the
original file is read.
Note that, special care must be taken for periodic parameters, Right
ascension and position angle (e.g., $p.a.=0 \deg$ is identical to
$p.a.=180 \deg$; thus, the mean of $p.a.=3 \deg$ and $p.a.=177 \deg$ 
is $p.a.=0 \deg$, not $p.a.=90 \deg$).

The adopted $\sigma_k$ are given in the following table:

\vspace{0.5cm}
\begin{tabular}{ll}
\hline
Parameter         & $\sigma_k$  \\
\hline
$\alpha_{2000}$   &  $30 \prime\prime$      \\
$\delta_{2000}$   &  $30 \prime\prime$      \\
$\log D$            &  $0.08$           \\
$\log R$            &  $0.08$           \\
$I_T$ &  $0.3$ magnitude  \\
$\beta$           &  $5/ \log R$ degree  \\
\hline
\end{tabular}
\vspace{0.5cm}

The standard error of the position angle is taken as a function of $logR$ because 
its meaning vanishes for face-on galaxies. 

\begin{figure}
\epsfxsize=8.5cm
\hbox{\epsfbox{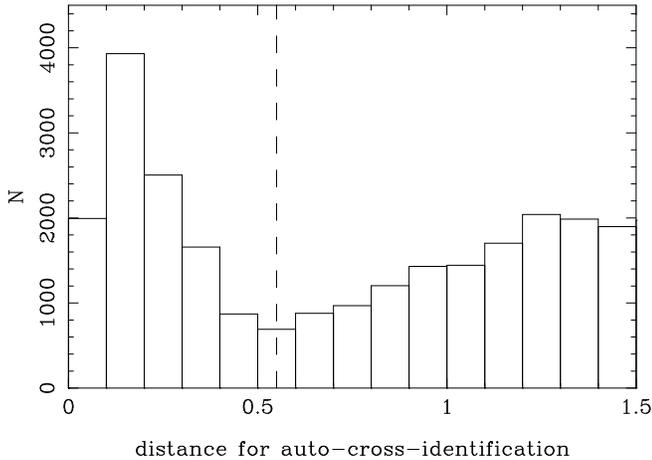}}
\caption{Distribution of ``distances'' $d_{ij}$ between two DENIS
objects ($i$ and $j$). This graph is used to chose $d_{limit}$
in the auto-cross-identification phase (see text).}
\label{fig7}
\end{figure}

Objects are merged for $d_{ij} < d_{limit}$,
$d_{limit}$ beeing chosen from the distribution of all distances
(Fig.\ref{fig7}). By its definition, $d_{ij}$ has the meaning of a Student's t-test
devided by $2$ (it is thus dimensionless). 
A $1\sigma$ criterion corresponds roughly to $d_{limit}=0.5$.
However, the value of $\sigma$ attached to each parameter is somewhat 
arbitrary, so is the 
definition of $d_{limit}$. We adopted $d_{limit}=0.55$. This
choice is guided by the minimum observed in the
histogram of $d_{ij}$ (Fig.\ref{fig7}).

During ``auto-cross-identification''
a provisional DENIS number (called RED) is given to each entry of the catalog.
When a galaxy appears several times in the catalog, each original set 
of measurements is identified with the same RED number. Each entry will be
cross-identified independently with LEDA galaxies, this will allow us to check
the reliability of the cross-identification.


\subsection{Cross-identification with LEDA galaxies}
From the previous step we get an intermediate catalog in which
each galaxy has a provisional number and all its astrophysical
parameters ($\alpha_{2000}$, $\delta_{2000}$, $logD_I$, $logR_I$, $I_T$ and $\beta$).
Each entry of this catalog must now be cross-identified with LEDA galaxies in order
to identify those already known. 

\begin{figure}
\epsfxsize=8.5cm
\hbox{\epsfbox{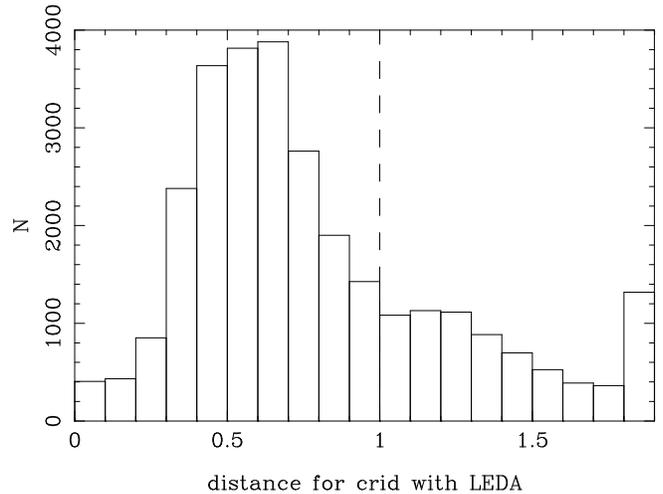}}
\caption{Distribution of ``distances'' $d_{ij}$ between DENIS and LEDA
galaxies ($i$ and $j$). This graph is used to chose $d_{limit}$
in the cross-identification phase (see text).}
\label{fig8}
\end{figure}

LEDA galaxies have similar parameters
$\alpha_{2000}$, $\delta_{2000}$, $logD_{25}$, $logR_{25}$, $B_T$ and $\beta$, but
diameters and magnitudes are defined in the photometric B-band.
The cross-identification is done by calculating the distance 
(in the mathematical sense, as defined by equation \ref{distance}) 
between a DENIS and a LEDA galaxy.
The coincidence is accepted if the distance is smaller than $d_{limit}$.
From a histogram of all distances between LEDA and DENIS measurements
(Fig.\ref{fig8}), we adopted $d_{limit}=1.0$  which corresponds to the first
minimum of $d_{ij}$-histogram (a pure Student's t-test would have given
$d_{limit}=1.29$ at a 0.01 probability level). 
This limit is voluntarily conservative (i.e., small)
because we prefer to miss a cross-identification than to merge two
distinct galaxies.

Because a given galaxy is cross-identified each time it appears in the catalog, 15945
were cross-identified several times with their original parameters. 
We reject 1881 galaxies identified with
more than one LEDA galaxy. The
final catalog contains now 36247 galaxies.

\subsection{Last cleaning}
Many objects were kept in the catalog despite the fact that they were labelled 
``undefined objects'' by the DA. They were kept because a known galaxy was very
close.
In the present release we removed these objects which
nature may be questionable without further inspection. Indeed, some
very faint galaxies in LEDA have only a few parameters, so that the cross-identification
is based mainly on coordinates. In some crowded field (clusters of galaxies or 
low galactic latitude) the accuracy of coordinates does not allow 
an identification secure enough. 
10696 such objects were then rejected, leaving 25551 galaxies.

Finally, we rejected galaxies with uncertain coordinates so that only coordinates
based on the GSC reference are used. So, 5291 galaxies are
rejected in the present version, leading to the final catalog of 20260 galaxies. 
These drastic rejections aim at maintaining a high quality 
level for this first catalog. In order to judge the quality more quantitatively
we will now compare with other sources of data.

\section{Comparison with Mathewson et al. samples}
\subsection{Magnitudes}
\begin{figure}
\epsfxsize=8.5cm
\hbox{\epsfbox{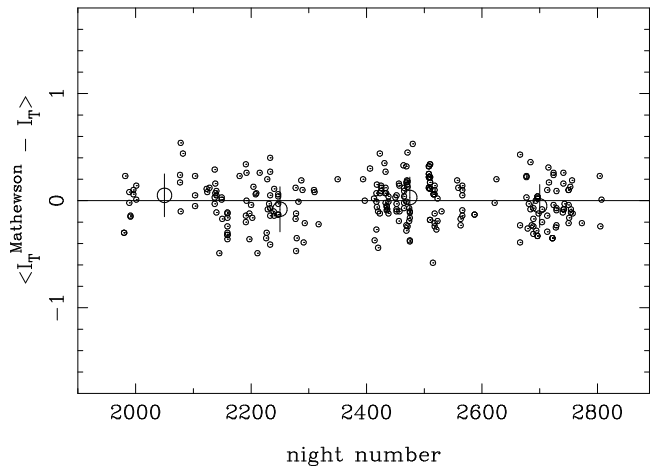}}
\caption{Zero-point variation obtained by comparison with Mathewson's
I-band photometry.}
\label{fig9}
\end{figure}

Magnitudes are calibrated by comparison with the measurements in I-band photometry
made by Mathewson et al. (1992, 1996) which gives access to 2441 galaxies. 
This comparison allows us to correct for a possible
variation of the zero-point. This variation has been explained by seasonal
variation of the mean temperature of the camera \footnote{the camera is now equiped with a 
regulated cooling system.}. Figure~\ref{fig9} shows such a variation
described by the parameter C:

\vspace{0.5cm}
\begin{tabular}{rc}
\hline
Night number &  C \\
\hline
1850 - 2150  & $ +0.05 $\\
2151 - 2350  & $ -0.08 $\\
2351 - 2600  & $ +0.03 $\\
2600 - 2800  & $ -0.05 $\\
\hline
\end{tabular}
\vspace{0.5cm}

Note that the ``night number'' is a logical number, not a real night number.
Each time the system is initialized the night number is incremented. This
explains that after one year of running survey there are 2500 logical nights.

An airmass correction is adopted  using a typical value
$\Delta I / \Delta sec \zeta = 0.04 \pm 0.02$. A check is made to control that
there is no airmass residual. The residual is smaller than 0.01 magnitude.
The DENIS I-magnitude is then:

\begin{eqnarray}
\lefteqn{I(DENIS) = }  \nonumber \\
 & & -2.5 log (\sum_{i=1}^{n} f_{ij}- f_{bg}) + 24.01 -0.04 sec \zeta + C
\end{eqnarray}

\begin{figure}
\epsfxsize=8.5cm
\hbox{\epsfbox{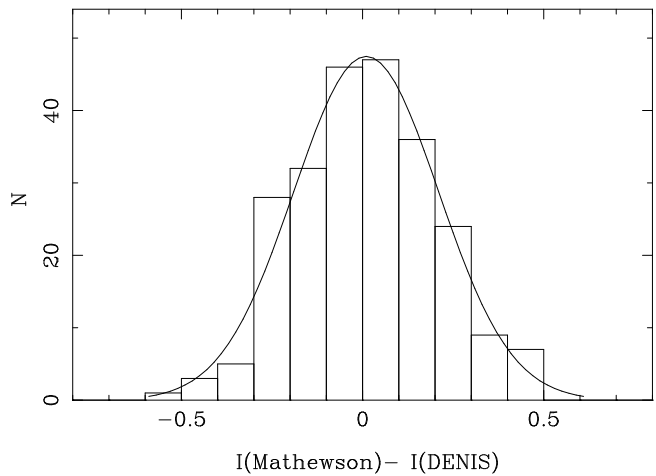}}
\caption{Zero-point distribution after a tiny seasonal correction}
\label{fig10}
\end{figure}

The zero-point distribution of $I(Mathewson) - I(DENIS)$
is Gaussian (Fig.\ref{fig10}) with a standard deviation of 0.2 magnitude. 
If we assume that the error is identical for both systems the 
mean error on DENIS extragalactic I-band magnitudes would be about 0.14 magnitude. 
Because the uncertainty on Mathewson et al. data is probably smaller, 
the uncertainty on our I-band magnitudes is about $\sim 0.18$ magnitude.

\begin{figure}
\epsfxsize=8.5cm
\hbox{\epsfbox{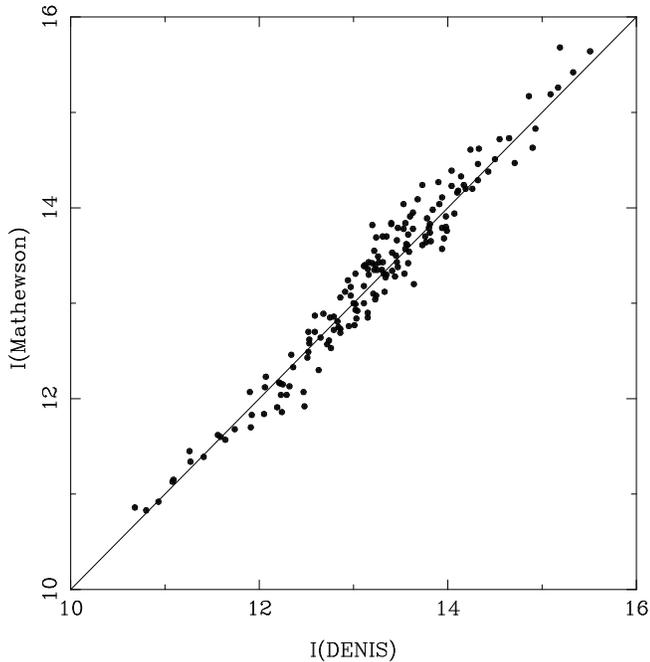}}
\caption{Comparison of extragalactic I-band
magnitudes from Mathewson et al. and from DENIS}
\label{fig11}
\end{figure}

In Fig.\ref{fig11}, the comparison between $I(Mathewson)$ and $I(DENIS)$
is shown for galaxies with secure identification and being neither ``multiple'' nor
``truncated''. The direct regression is: 
\begin{equation}
I(Math.) = 1.05 \pm 0.02 I(DENIS) -0.54 \pm 0.24
\end{equation}
with the following standard deviation, correlation coefficient and number of objects:
$\sigma=0.21$, $\rho=0.977 \pm 0.004$, $n=163$ after 11 rejections at $3\sigma$.

Stricto sensu, the slope is not significantly different from 1, and the zero-point
is not significantly different from 0. So, we are keeping the conservative
solution: $I(Math.) \equiv I(DENIS)$.
Among the 10 rejected galaxies, 8 can be explained by
localized poor photometric conditions (because they correspond to a loss of flux).
The measurements of corresponding nights will be counted with half weight. Two
nights were rejected (night 2475 and 2159) because they give rejections in the
comparison of different photometric parameters.

\subsection{Diameters ans Axis Ratios}
\begin{figure}
\epsfxsize=8.5cm
\hbox{\epsfbox{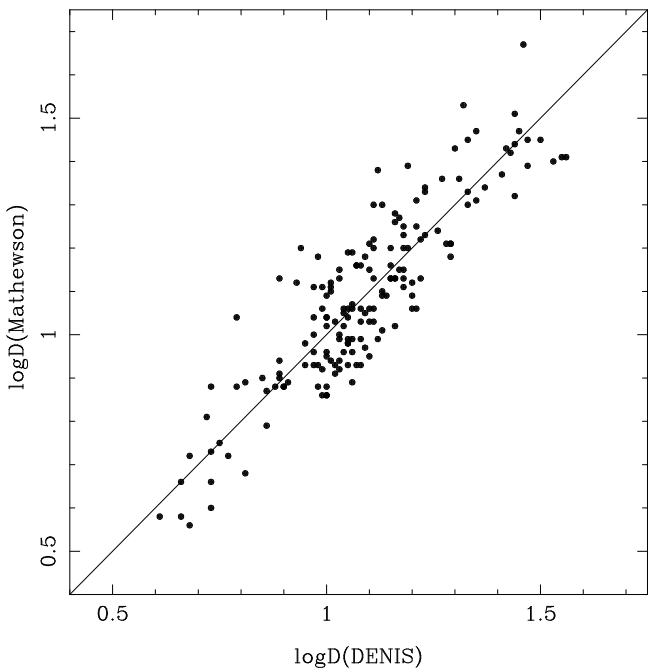}}
\caption{Comparison of I-band
isophotal diameters from Mathewson et al. and from DENIS.}
\label{fig12}
\end{figure}

\begin{figure}
\epsfxsize=8.5cm
\hbox{\epsfbox{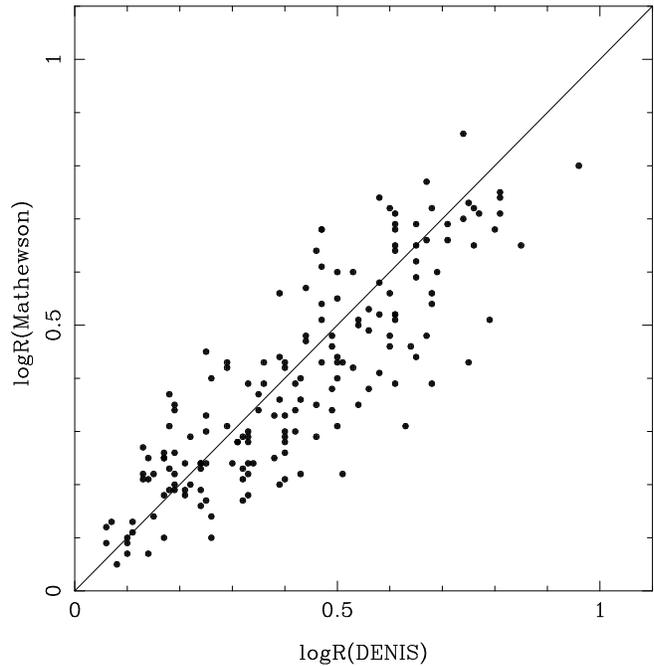}}
\caption{Comparison of I-band
axis ratios from Mathewson et al. and from DENIS}
\label{fig13}
\end{figure}

Diameters and axis ratios are also compared with those of Mathewson et al. samples.
These comparisons are given in Fig.\ref{fig12} and Fig.\ref{fig13}, respectively.
The direct regression are: 
\begin{eqnarray}
\lefteqn{logD(Math.) = } \nonumber \\
& & 0.96 \pm 0.04 logD(DENIS) +0.04 \pm 0.04
\end{eqnarray}
with $\sigma=0.10$, $\rho=0.88 \pm 0.02$, $n=170$ after 4 rejections at $3\sigma$.

For the axis ratio, it is better to force the intercept to be zero as generally
admitted (see de Vaucouleurs et al., 1976). This avoids to have negative axis
ratio after the application of the regression. The result is thus:

\begin{equation}
logR(Math.) = 0.94 \pm 0.04 logR(DENIS) 
\end{equation}
with $\sigma=0.10$, $\rho=0.85 \pm 0.02$, $n=172$ after 2 rejections at $3\sigma$.

None of these regressions is significantly different from the absolute identity.
So, we will keep: $logD(Math.) \equiv logD(DENIS)$ and $logR(Math.) \equiv logR(DENIS)$.
The standard deviations are 0.10 for both $logD$ and $logR$. Again, if
we assume the same error on both systems the error on $logD$ and $logR$
is about 0.07. 

\section{Comparison with LEDA galaxies}
\subsection{Equatorial coordinates}
\begin{figure}
\epsfxsize=8.5cm
\hbox{\epsfbox{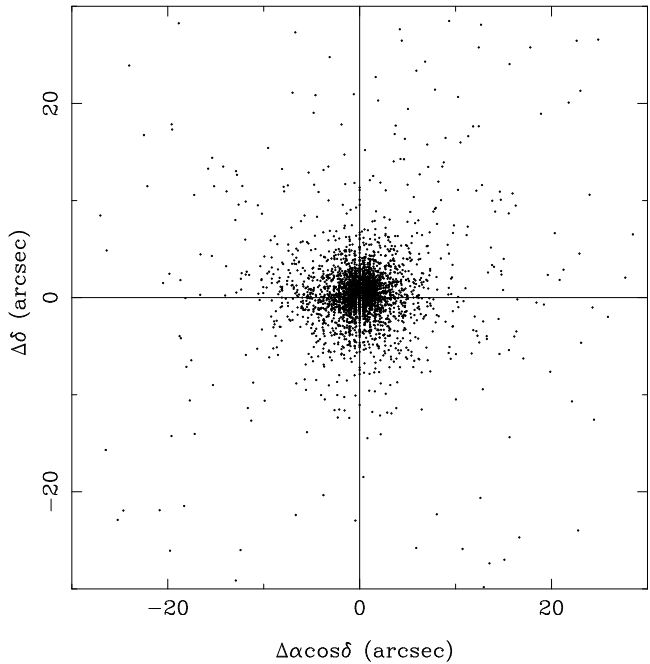}}
\caption{Comparison of coordinates from LEDA and DENIS.}
\label{fig14a}
\end{figure}

The coordinates are compared with coordinates given in LEDA. Only
coordinates known as 'accurate' in LEDA (i.e.with a standard deviation
less than $10$ arcseconds) are used. 
The plot of $\Delta \delta$ and $\Delta \alpha.cos \delta$ in Figure \ref{fig14a} shows
that there is no systematic distorsion ($\Delta$  means $(DENIS)-(LEDA)$).
The standard deviation is $6.5$ arcseconds and $6.7$ arcseconds for $\alpha.cos \delta$ and
$\delta$, respectively. Assuming an error of the same amplitude in LEDA and DENIS
coordinates gives an uncertainty of $6.6/\sqrt{2}$ arcseconds for DENIS right 
ascension and declination, i.e. an uncertainty of $6.6$ arcseconds for the position of
a galaxy.

In fact, the plot of $\Delta \alpha cos \delta$ vs. $\Delta \delta$ exhibited two abnormal zones with
a systematic shift of about 30 or 40 arcseconds. 
This problem appeared near the zenith. 
Thus, only objects with coordinates based on the GSC stars are kept in the present
version.  In the final catalog, the coordinates will be obtained through a full astrometric
solution (mosaicing frames along each strip and with adjacent ones) and this problem
will be solved without rejecting galaxies. 

\subsection{Position angle}
\begin{figure}
\epsfxsize=8.5cm
\hbox{\epsfbox{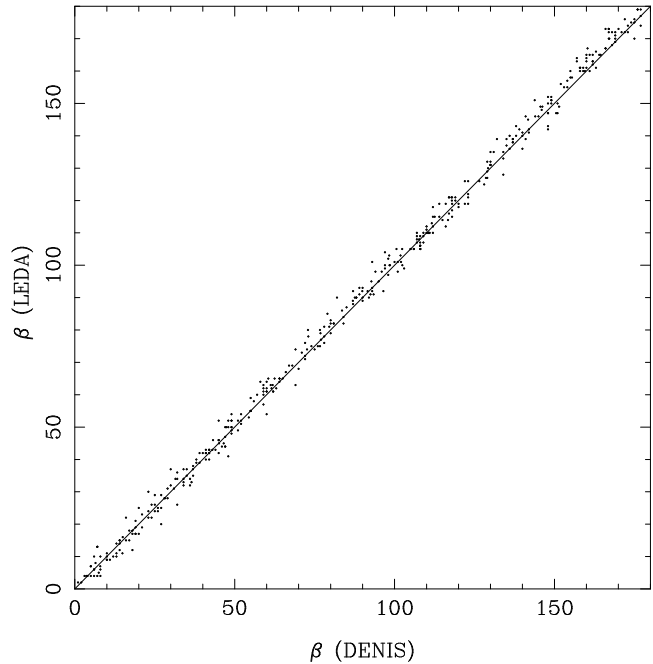}}
\caption{Comparison of 
position angles from LEDA and DENIS.}
\label{fig15}
\end{figure}

Position angle is important for studies on the orientation of galaxies, but also
for identification. However, for nearly face on galaxies, it becomes very uncertain.
The comparison is made only for galaxies with $logR(DENIS) > 0.5$. The result is given 
in (Fig.\ref{fig15}). 
The direct regression gives:
\begin{equation}
\beta (LEDA) = 1.009 \pm 0.003 \beta (DENIS) +0.07 \pm 0.26
\end{equation}
with $\sigma=2.65$, $\rho=0.9987 \pm 0.0001$, $n=408$ after 21 rejections at $3\sigma$.
Among the 21 rejections, only 2 correspond to real discrepancies.

The uncertainty on the measurement of 
$\beta$ is $2.7 \deg.$. This excellent agreement of position angles validates the
reliability of our cross-identification.

\subsection{Photometric morphological type code}
It is interesting to obtain at least a very crude estimate of the morphological 
type code. This is particularly important when we plan to start 
a HI follow-up  for which it is compulsory to avoid elliptical galaxies. 
The log of the standard deviation of pixel intensities is significantly correlated
with the morphological type code extracted from LEDA. The relation is almost linear.
However, it appears that the solution depends on the size of 
the galaxy (all very small objects appear identical). 
The slope and the intercept are found linearly correlated with the
log of the number of pixels.  
In the scale between -5 and 10 defined in the Second Reference 
Catalog of Bright Galaxies, 
the photometric morphological type code is calculated as:

\begin{equation}
T(DENIS) = A \,\log(\sigma(f_{ij})) + B
\end{equation}
Where $\sigma(f_{ij})$ is the standard deviation of flux of all pixels of the
matrix representing the galaxy :
\begin{equation}
\sigma(f_{ij}))=\sqrt{\frac{\sum_{i,j} (f_{ij} - \bar{f})^2}{(n-1)}}
\end{equation}
with 
\begin{equation}
A= \frac{1}{-0.0187 \,\log n -0.0183}
\end{equation}
\begin{equation}
B= \frac{1}{ 0.2774 \,\log n +1.7846}
\end{equation}
where $n$ is the number of pixels.

\begin{figure}
\epsfxsize=8.5cm
\hbox{\epsfbox{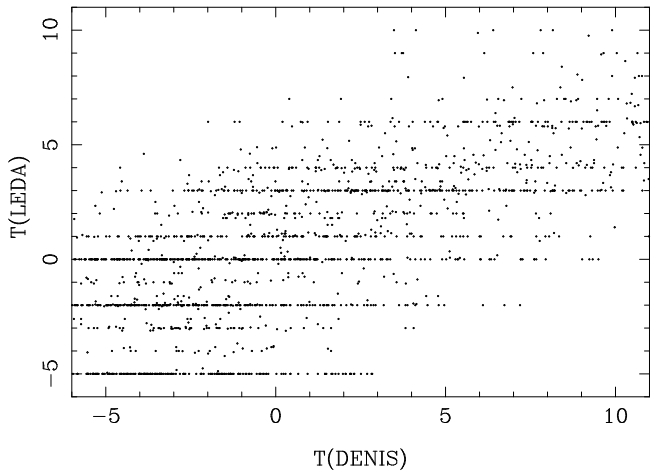}}
\caption{Comparison 
of morphological type codes from LEDA and from the I-band photometry.}
\label{fig16}
\end{figure}

The comparison of codes $T(DENIS)$ and $T(LEDA)$ is given in Figure~\ref{fig16}.
The correlation is clearly significant (the correlation coefficient is $\rho = 0.69\pm0.01$) but the
standard deviation is large ($\sigma(T(DENIS))=2.5$).
This allows us to classify galaxies into 'Early', 'Intermediate'
and 'Late' types, with no finer subdivision.
In the catalog, photometric morphological type smaller than $-5$ or greater than $10$
will be set to $-5$ or $10$, respectively.

\section{The catalog}
\subsection{Description of the catalog}
The catalog of galaxies which is described here is the first of a series which will be released
during the progression of the survey. At the end of the survey a deeper and complete
catalog will be produced at PDAC. The present one, is necessarily limited to the
first observations (one year).

The catalog contains 20260 galaxies. Among them 14518 are new galaxies 
while 5742 are galaxies already present in LEDA. 

Each galaxy is numbered with a provisional internal number (RED, for Rapid Extraction from DENIS), 
and an identifier from the LEDA database (PGC/LEDA)
 \footnote{connection: \\ http://www-obs.univ-lyon1.fr/leda \\ or \\ {\bf telnet:} lmc.univ-lyon1.fr {\bf login:}leda}. 
Galaxies are also identified with an 
alternate name taken in the following catalogues:
NGC, IC (Dreyer, 1988), UGC (Nilson, 1973), ESO (Lauberts, 1982), MCG (Vorontsov-Velyaminov
et al., 1962-1974), CGCG (Zwicky et al., 1961-1968), IIZW, IIIZW, VIII from the
catalog of compact and eruptive galaxies (Zwicky, 1971), IISZ (extension of the same list by
Rodgers et al., 1978, FAIR (Fairall, 1977-1988), HICK (Hickson, 1989), KUG (Takase and Miyauchi-Isobe, 1984-1989), IRAS (IRAS), MK (Markarian, 1971-1977 ), UM (University of Michigan list, 
MacAlpine et al., 1981).
The detailed references to these catalogs are given in Paturel et al. (1989). 
Two additional catalogs are included: Saito et al. (1990-1991) and Dressler (1980). 
The notations for corresponding galaxy names are SAIT
(e.g. SAIT 69-1, for object 69 in list 1 of Saito et al.) and DRCG 
(e.g., DRCG 39-41, for galaxy 41 in cluster 39, the numbering of clusters is
made according to the order in the original publication).

For each galaxy, the catalog gives the weighted mean parameters obtained as described in the
previous section. Actual mean errors $\sigma$ are calculated according to Paturel et al. (1996)
using individual mean errors deduced in previous section. For I-band magnitudes
the mean error is taken as a function of the magnitude itself. An estimate made
from Fig.~\ref{fig11} gives:$\sigma (I) = 0.05B_T-0.51$. The weights used for calculating
the weighted means are the inverse square of actual mean errors. For nights suspected to
be of poor photometric quality the actual mean error is divided by 3\footnote{ These
nights were detected by a $3 \sigma$ rejection rule, so one can admit that their mean error
is three times the typical mean error.}. Further, when the quality of
the matrix is uncertain ('peculiar', 'multiple' or 'truncated'), the weight
is divided by 2. This correcting coefficient is deduced from the comparison of
I magnitudes with poor quality objects.

The quality of each individual matrix is coded as follows:
'Normal' galaxies are coded as 1, 'Peculiar' galaxies as 10, 'Multiple' as 100
and 'Truncated' as 1000. This code is added for each measurement. So, the resulting
code gives immediately the number of independent measurements (sum of digits) and
the quality of each of them. For instance, the code 1002 means that there are
one truncated matrix and two normal ones (i.e., 3 independent measurements).
'Truncated' means an overestimated magnitude, on the contrary,
'Multiple' means that the magnitude of the considered galaxy is underestimated.

The different columns are the following:
\begin{itemize}
\item {\bf Column 1:}Provisional DENIS identifier. This identifier is just an internal
number given for easy identification. This number will be replaced by an appropriate
final DENIS number.
\item {\bf Column 2:}LEDA identifier. This number allows rapid retrieval in the LEDA database.
This number is identical to the PGC number for LEDA number less than 73197.
\item {\bf Column 3:}Alternate name according to a hierarchy: NGC, IC, ESO, UGC, MCG (see
text above).
\item {\bf Column 4:}Equatorial coordinates for epoch 2000, in hours, minutes, seconds and
tenths for the Right Ascension and in degrees, arcminutes and arcseconds for
the declination. The actual mean error on the position of the galaxy is given in the same column, 
next line in arcseconds.
\item {\bf Column 5:} Total I-band magnitude $I(DENIS)$ with its mean error on the second line.
\item {\bf Column 6:} Isophotal diameter in log scale $logD(DENIS)$, where $D$ is in 
$0.1$ arcminute following the notation of de Vaucouleurs et al. (RC1, RC2). The actual mean error is
given on the second line.
\item {\bf Column 7:} Isophotal axis ratio in log scale $logR(DENIS)$, where $R$ is  the
ratio of the major to the minor axis. The actual mean error is given on the second line.
\item {\bf Column 8:} Position angle $\beta (DENIS)$ in degrees, measured from North Eastwards. 
Its value lays between $0$ and $180$ degrees (excluding $180$). For $logR(DENIS) < 0.5$ the
meaning is poor. This is reflected by the actual mean error given on the second line.
\item {\bf Column 9:} Photometric morphological type code $T(DENIS)$. 
The coding is given according to RC2
(i.e. $T$ from $-5$ to $10$ for Elliptical to Irregular galaxies).
The mean error is given on the second line.
\item {\bf Column 10:} Quality code of the object (1000=truncated object, 100=multiple object,
10=peculiar object, 1=normal galaxy). The sum of digits gives the number of independent
measurements.
\end{itemize}

A sample page is given at the end of the paper. The catalog is available in electronic form
at PDAC and Lyons Observatory.

\subsection{Distribution on the sky}
\begin{figure*}
\epsfxsize=18.0cm
\hbox{\epsfbox{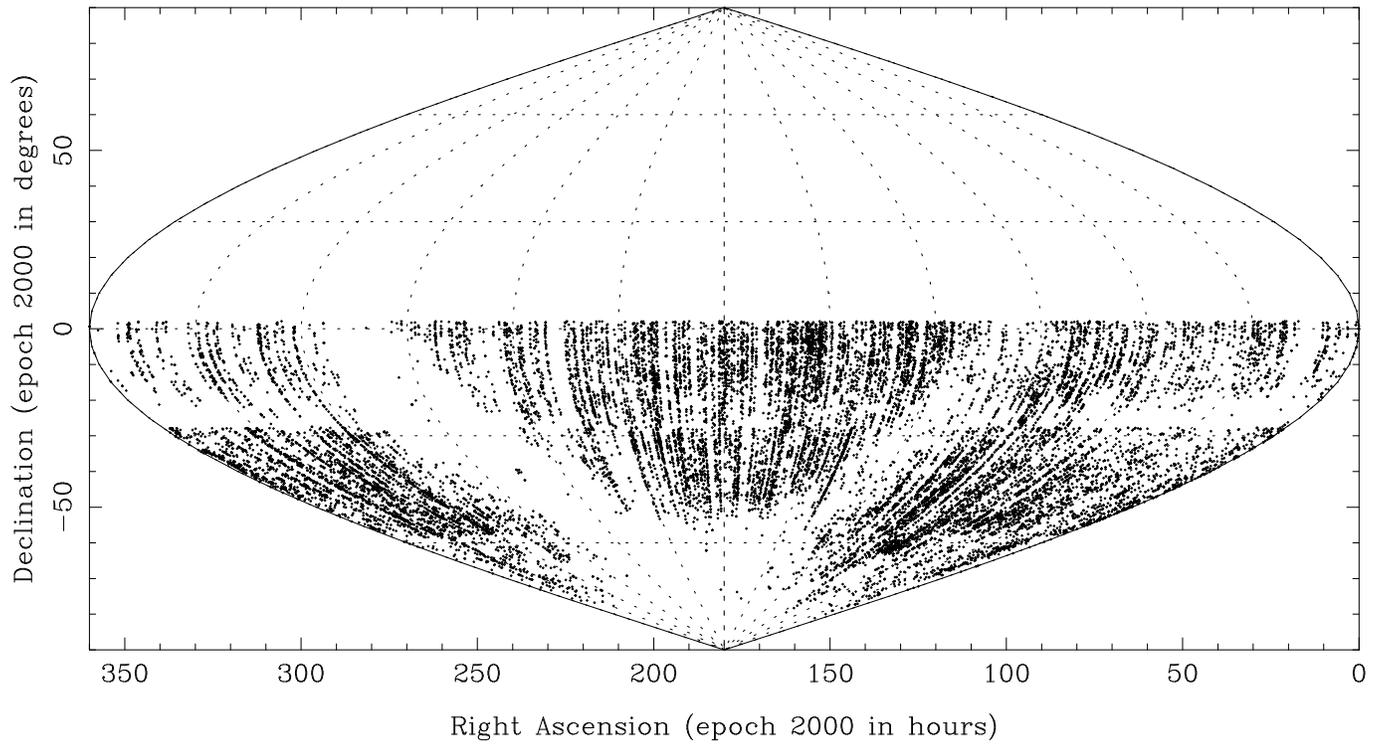}}
\caption{Distribution of the 20260 galaxies of the present DENIS catalog
presented with a Flamsteed's equal area projection in equatorial
(2000) coordinates.}
\label{fig17}
\end{figure*}

The distribution on the sky is given in equatorial coordinates for the epoch 2000 
(Figure \ref{fig17}).
The strips are clearly visible. The zone called 'Equatorial' ($\delta$ between $+2\deg$ and $-28\deg$)
is not so well covered because observations are avoided in this zone when Moon is bright or when
the wind is strong. Further, many frames are rejected in this zone 
at the end of the strip (near $-28\deg$) because of the shift in header-coordinates. 
No attempt is made to reach low galactic latitudes. This is done independently in J and K bands.

\subsection{Completeness limit}
\begin{figure}
\epsfxsize=8.5cm
\hbox{\epsfbox{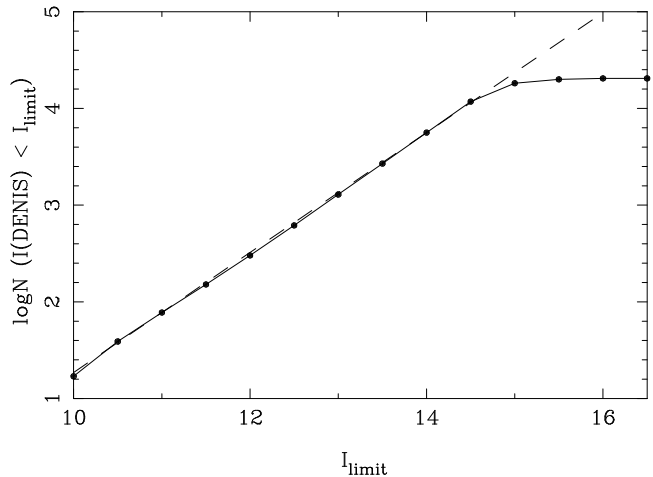}}
\caption{Completeness curve. The completeness curve is quite linear up to the
completeness limit $I_{limit}= 14.5$. The slope ($0.62 \pm 0.02$, dashed
line) is very nearly the one expected for a homogeneous distribution 
of galaxies (0.6).}
\label{fig19}
\end{figure}

If we assume a homogeneous distribution of galaxies ($N \propto r^{3}$), the
plot of the number of galaxies (in log scale, i.e. $logN$) brighter than a given 
magnitude limit $I_{limit}$
versus this magnitude limit should be linear with a slope of $0.6$.
This plot (Figure \ref{fig19}) is quite linear up to $I_{limit}= 14.5$
with a slope of $0.62 \pm 0.02$.
The sense of this completeness limit must be explained. It means that in surveyed
directions all galaxies brighter than 14.5 are detected. Because the sampling is
made randomly and $1/4$ of the survey
is presented here, the completeness limit in apparent magnitude 
of this first DENIS catalog is 
$I_{limit}= 14.5$.

\acknowledgements{
This work would have been impossible without the collaboration of 
Marthinet M.C., Petit C., Provost L., Gallet F., Garnier R., Rousseau J.
They are fully associated with this work.
The DENIS team is warmly thanked for making this work possible and in
particular the Operations team at la Silla. We thank also G. Mamon for
careful reading of the article.\\
The DENIS program is partly funded by the European Commission through
{\it SCIENCE} and {\it Human Capital and Mobility} grants. It is also
supported in France by INSU, the Education Ministry and CNRS, in
Germany by the Land of Baden-W\"urtenberg, in Spain by DGICYT, in
Italy by CNR, in Austria by the Fonds zur F\"orderung der Wissenschaft
und Forschung, in Brazil by FAPESP.
}


\newpage
\begin{table*}
\begin{verbatim}
TABLE 1: Sample page of the present catalog.
----------------------------------------------------------------------------------------------
DENIS-RED LEDA/PGC  Alternate Name  R.A. DEC.(2000)      I     logD   logR  p.a.    T     code 
   (1)       (2)          (3)           (4)             (5)     (6)    (7)  (8)    (9)    (10) 
==============================================================================================
000142    000087   DRCG  39- 41    J000109.8-503505    14.82   .718   .306   42.9   0.       2
                                                  5      .17   .061   .057    5.5   2.
000149    143168                   J000110.8-404542    12.97   .873   .083   17.0  -4.       1
                                                  6      .14   .070   .070   19.0   3.
000150    143169                   J000111.7-384507    15.45   .733   .486  102.0   6.       1
                                                  6      .26   .070   .070    3.9   3.
000153    124807                   J000112.3-391354    14.99   .648   .188   95.9  -1.       2
                                                  4      .21   .052   .080   21.5   3.
000159    143172                   J000115.9-730851    15.28   .683   .417  120.0  -2.       1
                                                  6      .36   .099   .099    6.4   4.
000160    000099   ESO  149- 12    J000117.5-530033    13.54  1.138   .701   36.9   4.       2
                                                  5      .12   .050   .054    2.2   2.
000165    143175                   J000119.7-523551    14.29   .828   .389   14.1  10.       2
                                                  5      .15   .067   .051    4.0   2.
000168    143177                   J000120.4-404721    14.70   .713   .236   44.0   1.       1
                                                  6      .23   .070   .070    7.7   3.
000170    143179                   J000121.0-723533    14.97   .683   .181  134.0   0.       1
                                                  6      .24   .070   .070    9.9   3.
000174    130913                   J000131.1-404912    13.99   .873   .208  114.0  10.       1
                                                  6      .19   .070   .070    8.7   3.
000182    124990                   J000138.5-435949    14.51   .663   .000  153.0  -2.       1
                                                  6      .22   .070   .070   90.9   3.
000183    124811                   J000139.9-383834    15.59   .543   .264   46.0   2.       1
                                                  6      .27   .070   .070    7.0   3.
000226    143200                   J000200.1-515740    14.64   .713   .181   39.0   0.       1
                                                  6      .22   .070   .070    9.9   3.
000253    143209                   J000218.6+015033    14.53   .683   .181  134.0  -4.       1
                                                  6      .22   .070   .070    9.9   3.
000272    143215                   J000233.8-111705    14.31   .903   .458   78.0   3.       1
                                                  6      .21   .070   .070    4.2   3.
000289    000205   UGC       5     J000305.7-015449    12.18  1.248   .354   51.5   5.       2
                                                  5      .07   .052   .069    4.2   2.
000292    143223                   J000307.6-160643    14.64   .723   .153   90.0   3.       1
                                                  6      .22   .070   .070   11.4   3.
000293    143224                   J000308.6-170300    14.55   .683   .125   83.0  -2.       1
                                                  6      .22   .070   .070   13.6   3.
000294    143225                   J000309.0-195221    15.00   .703   .431   19.0  -5.       1
                                                  6      .24   .070   .070    4.4   3.
000296    000211                   J000310.5-544457    12.23  1.135   .167   58.1  -4.    1001
                                                  6      .10   .071   .078    9.9   3.
000299    143226                   J000313.5-512806    14.78   .743   .417  142.0   0.       1
                                                  6      .23   .070   .070    4.6   3.
000303    143228                   J000318.3-131618    14.16   .923   .417   83.0  10.       1
                                                  6      .20   .070   .070    4.6   3.
000307    143229                   J000320.4-394823    14.96   .783   .486  131.0   3.       1
                                                  6      .24   .070   .070    3.9   3.
000309    000224   FAIR    627     J000321.3-500448    13.47   .933   .347  110.2  -4.       2
                                                  5      .12   .053   .109    3.9   2.
000310    073217                   J000321.4-543338    12.81  1.023   .125  145.0  -2.     100
                                                 20      .39   .210   .210   40.8   9.
000311    143230                   J000321.6-190604    14.74   .733   .306  146.0  -2.       1
                                                  6      .23   .070   .070    6.1   3.
000313    143232                   J000322.4-434615    14.73   .693   .208  165.0  -2.       1
                                                  6      .23   .070   .070    8.7   3.
==============================================================================================
\end{verbatim}
\end{table*}
\end{document}